\documentclass[letterpaper, 10 pt, conference]{ieeeconf}  

\usepackage[utf8]{inputenc}
\usepackage{algorithmic}
\usepackage{textcomp}
\usepackage{cite}
\usepackage{cancel}
\usepackage{graphicx}
\PassOptionsToPackage{final}{graphicx}

\usepackage{amsmath}
\usepackage{amssymb}

\usepackage{amsthm}
\usepackage[version=4]{mhchem}
\usepackage{siunitx}
\usepackage{longtable,tabularx}
\usepackage{mathtools}
\usepackage{easyReview}
\usepackage{cleveref}
\usepackage{caption}
\usepackage{subcaption}
\usepackage{amsthm,thmtools,xcolor}
\definecolor{nblue}{rgb}{0,0.263,0.576}
\definecolor{mblue}{rgb}{0.075,0.541,0.855}
\newtheorem{defn}{Definition}
\newtheorem{lemma}{Lemma}
\newtheorem{theo}{Theorem}
\newtheorem{prop}{Proposition}

\newcommand{\trp}{\scriptstyle\top}

\newcommand{\R}{\mathbb{R}}

\IEEEoverridecommandlockouts                              
\title{\LARGE \bf
Set input-to-state stability under input delays for nonlinear systems with disturbances}

\author{Pallavi Sinha, Irinel-Constantin Mor\u{a}rescu, Sukumar Srikant 
\thanks{P. Sinha  {\tt\small (pallavisinha33@gmail.com)} and S. Srikant {\tt\small (srikant.sukumar@iitb.ac.in)} are with the Systems and Control Engineering, Indian Institute of Technology Bombay, Powai, Mumbai, 400076, India. Irinel-Constantin Mor\u{a}rescu {\tt\small (constantin.morarescu@univ-lorraine.fr)} is with Universite de Lorraine, CNRS, CRAN, F-54000 Nancy, France.}}

\begin{document}

\maketitle
\thispagestyle{empty}
\pagestyle{empty}

\begin{abstract}
The study proposes new results on the set input-to-state stability (ISS) subject to a small input time delay for compact, invariant sets that contains the origin. First, using the nonlinear small-gain theory, we prove a Razumikhin-type theorem that ensures ISS for sets in the setup of functional differential equations (FDEs) with disturbances. Next we show how this theorem can be used to ensure set ISS for nonlinear systems with input delays and disturbances. In comparison to the existing research on set ISS robustness with respect to small time delays at the input, our results are rather broad, retaining the ISS gain and without any constraints on time delayed states. The effectiveness of the method is illustrated through two case-studies addressing  set stability for classes of nonlinear oscillators of practical interest.
\end{abstract}
\section{Introduction}
Time delays are ubiquitous in modern control applications. In many of these applications the source of time delays is the presence of communication networks between sensors, plant and controllers. Delayed states can also be found in a variety of engineering systems, that don't involve communication networks such as advanced defence systems, rolling mills, automotive engines, laser models, chemical and biological reactors \cite{hale2013introduction,fridman2014introduction,kolmanovskii2013introduction}.
\newline\newline 
It is well established that time delay systems (TDS) are a class of infinite-dimensional FDEs. The methods for stability analysis of TDS can be structured in frequency-domain \cite{olgac2002exact},\cite{moruarescu2007stability} and time-domain ones. The time-domain methods extend Lyapunov approach and they are based either on Lyapunov-Krasovskii functionals \cite{Krasovskii1963stability} or on the Lyapunov-Razumikhin ones \cite{Razumikhin1956500}. Detailed descriptions of these methods and their applications can be found in \cite{michiels2007stability},\cite{niculescu2001delay}. Stability analysis of linear TDS mainly based on linear matrix inequalities (LMI) is addressed in \cite{Fridman2002improved}, \cite{Fridman2004robust} and references therein. While the literature on the analysis and the control design of linear TDS is quite complete, several issues remain only partially addressed for nonlinear TDS. \\

An important topic in control research is the stability under input perturbations. The robustness of a class of nonlinear systems w.r.t. time delays in the input has been addressed in \cite{fan2006delay}. In \cite{mazenc2013robustness} the authors studied the robustness of nonlinear systems w.r.t. small time delays and sampling by using an explicit Lyapunov-Krasovskii functional. While the Lyapunov-Krasovskii technique is a logical extension of the Lyapunov approach, the choice of a Lyapunov-Krasovskii functional, remain a complicated problem. In many nonlinear scenarios it is easier to find a Lyapunov-Razumikhin function (see for instance [26] and [27]). In other words, Lyapunov-Razumikhin technique has a certain advantage when one deals with real application of TDS stability tools.\\

The concept of ISS was first put forth as a theoretical foundation for the investigation of robustness issues in \cite{sontag1989smooth}, and many researchers have since used it in a variety of contexts, from robust control \cite{tsinias1989sufficient} to highly nonlinear small-gain theorems \cite{jiang1994small}. For FDEs with disturbances, ISS is guaranteed by Razumikhin-type theorems, which were proposed in \cite{teel1998connections}. The author in \cite{teel1998connections} also demonstrated that ISS is robust to small input time delays. In \cite{chaillet2022iss} the authors present a nice survey on ISS for TDS based on the Lyapunov-Krasovskii approach. \\

The ISS property, along with fundamental stability notions, was initially defined with regard to a specific equilibrium state of interest. Later, systems like boolean control networks \cite{guo2015set} or a group of interconnected systems that converge to or can be stabilized to a subset of the state space came into the picture. In \cite{guo2015set}, both concepts are referred to as set stability and set stabilization, respectively. A group of locally interconnected physical, biological, chemical, technological, and social systems can be synchronized using set stability (see for instance \cite{pscopsyncElsys}). The authors in \cite{shi2009global} consider group of continuous time agents and investigate set stability to achieve state agreement in the presence of network leaders who have apriori knowledge of the set of convergence. Several types of input-to-output stability as well as other stability problems including incremental stability, robust consensus/synchronization, ISS of time-varying systems, are covered under a more comprehensive definition of ISS with relation to a closed set \cite{noroozi2017nonconservative}. Seminal works on set ISS include \cite{lin1996smooth}, \cite{lin1995input}. The various set ISS ideas are proved for a compact invariant set in particular in \cite{sontag1995characterizations}. In \cite{shi2011connectivity} the authors have seen set ISS for set tracking when non-stationary network leaders define the desired set of accumulation over time. Set-stability of multi stable systems, robustness of set ISS to additive time delays in input using Lyapunov-Razumikhin functions, and their application in droop-controlled micro grids are discussed in \cite{efimov2016robustness}.
\newline\newline
 The various concepts of stability of invariant sets of FDEs are introduced in \cite{bernfeld2003stability}. The authors in \cite{yang2015stability} and \cite{efimov2020estimation} provide a Razumikhin-type version of set ISS for FDEs in their works. To the best of our knowledge, the literature lacks a proof for the set ISS Razumikhin-type theorems for TDS using small gain argument. These findings are rather useful to demonstrate that set ISS for nonlinear finite dimensional control systems is robust to small input time delays in feedback. To the best of the authors knowledge the only rigorous result guaranteeing set ISS for a class of nonlinear systems in presence of small time delay in the input is \cite{efimov2016robustness}. The particularity in \cite{efimov2016robustness} is that the time delayed states affect the input in an additive manner.\newline\newline
The article's novel contributions in comparison to the body of literature are listed below.
 \begin{enumerate}
     \item  We provide detailed proofs based on small gain argument for different Razumikhin versions of set ISS for FDEs.
     \item We propose a more general setup to address this gap and demonstrate the set ISS of nonlinear system's robustness to input time delays without any restrictions on time delayed states. This encompasses the results in \cite{efimov2016robustness} on preservation of set ISS in presence of small time delay at the input, where the input is dependent on a time delayed state as an additive effect.
     \item In comparison to \cite{efimov2016robustness}, the ISS gains are unchanged for systems with and without time delay.
     \item Using two illustrative examples, it is demonstrated that our method is effective without the presence of any restrictive requirement on time delayed states. These examples also highlight that the constraints in \cite{efimov2016robustness} are not easy to hold. 
 \end{enumerate}
 \vspace{0.2cm}
The remainder of the paper is set out as follows. The relevant notations, definitions and theorems are initially provided in Section \ref{prelim}. The proofs of the set-versions of the Razumikhin-type theorem, input-to-state stabilizibility, and the major finding of the paper on robustness of the set ISS to small time delays at input, are all illustrated in the Section \ref{mainres}. We also compare the findings of our robustness study to those in the article \cite{efimov2016robustness} which is well illustrated using examples in Section \ref{examplesection}. The future road map is presented in Section \ref{conclusion} of our study.
\section{Preliminaries}\label{prelim}
The goal of this section is to introduce some results which help us prove the robustness of set ISS to small time delays at input.\newline\newline
A mapping $\gamma:\mathbb{R}_{\geqslant 0} \to \mathbb{R}_{\geqslant 0}$ is said to belong to class-$\mathcal{G}$ if it is continuous, zero at zero and non-decreasing; it is of class-$\mathcal{K}$ if it is of class-$\mathcal{G}$ and strictly increasing; it is of class-$\mathcal{K}_{\infty}$ if it is of class-$\mathcal{K}$ and unbounded. A continuous function $\beta:\mathbb{R}_{\geqslant 0} \times\mathbb{R}_{\geqslant 0} \to \mathbb{R}_{\geqslant 0}$ is said to be of class-$\mathcal{KL}$ if $\beta(\cdot,t)$ is of class-$\mathcal{K}$ for each $t \geqslant 0$ and $\beta(s,\cdot)$ is monotonically decreasing to zero for each $s>0$. Throughout the paper, the following standard notations are used: set $\mathcal{C}=\mathcal{C}([-\delta,0],\mathbb{R}^n)$ represent continuous functions mapping the interval $[-\delta,0]$ into $\mathbb{R}^n$, with the topology of uniform convergence and some $\delta>0$. At the time $t$ the state vector is $x(t)$ and for the sake of simplicity, when no confusion can arise, we will denote it just by $x$. The state function $x_t$ corresponds to the past time-interval $[t-\delta, t]$, where $x_t(\theta)=x(t+\theta)$, where $\theta\in[-\delta,0]$.
\subsection{Set-stability for systems without time delay}
Consider nonlinear systems of the following form
\begin{align}\label{odeiss}
    \dot{x}(t)=f(x(t),u(t)),
\end{align}
where $x(t)\in \mathbb{R}^n$, $u(t)\in \mathbb{R}^m$, $\forall t \geqslant 0$, and the map $f:\mathbb{R}^n\times \mathbb{R}^m \to \mathbb{R}^n$ is smooth. The input $u$ is measurable, locally essentially bounded function of the type $u:\mathbb{R}_{\geqslant 0} \to \mathbb{R}^m$. The space of such functions is denoted by $\mathcal{L}^m_{\infty}$ with the norm $\|u_{[t_0,\infty)}\|\triangleq\text{ess} \sup\{u(t):t\geqslant t_0 \geqslant 0\}$. 
We use $\|u\|=|u_{[0,\infty)}|$ and let $\|u_{[0,t]}\|$ be the signal norm over the truncated interval $[0,t]$. For each initial state $x_0=x(0)\in \mathbb{R}^n$ and each $u \in \mathcal{L}^m_{\infty}$, let $x(t,x_0,u)$ denote the solution of (\ref{odeiss}) at time $t$. If there is no ambiguity from the context, the solution is simply written as $x(t)$. 
\newline\newline   For a nonempty closed set $A\in \mathbb{R}^n$ we have, $|y|_A\triangleq \inf\{d(y,z):z\in A\}$ where $d(y,z)=|y-z|$. 
\begin{defn}[Finite escape time\cite{teel2002integral}]
The system (\ref{odeiss}) is finite escape-time detectable through $|\cdot|_A$  if, a solution’s maximal interval of existence is bounded, i.e., $x(t)$ is defined exclusively on $[0, T)$ with $T$ finite, then $\displaystyle{\lim_{t\to T}} |x(t)|_A = \infty$.
\end{defn}
 \begin{defn}[Invariant set w.r.t ODE\cite{teel2002integral}]
 For the associated "zero-input" system
 \begin{align*}
     \dot{x}(t)=f(x(t),0),
 \end{align*}
set $A$ is said to be a $0$-invariant set if it holds that for any $x_0 \in A$, $x(t,x_0,0) \in A$ for all $t \geqslant 0$.
\end{defn}

The rest of the document will simply refer to the $0$-invariant set as the invariant set.
 \begin{defn}[Set ISS\cite{teel2002integral}]
 If there exists $\beta\in \mathcal{KL}$ and $\gamma \in \mathcal{K}$ such that for each $u \in \mathcal{L}_{\infty}^m$ and all initial states $x_0$, the solution $x(t)$ is defined  for all $t \geqslant 0$ and satisfies
 \begin{align*}
     |x(t)|_A\leqslant  \beta(|x_0|_A,t)+\gamma(\|u_{[0,t]}\|),
 \end{align*}
  for each $t \geqslant 0$. Then the system (\ref{odeiss}) is ISS with respect to a closed, invariant set $A$ 
 \end{defn}
\begin{defn}[ISS-Lyapunov function w.r.t. sets\cite{teel2002integral}]\label{setisslyap}
A smooth ISS-Lyapunov function for (\ref{odeiss}) with respect to the closed set $A$ is a smooth function $V:\mathbb{R}^n\to \mathbb{R}_{\geqslant 0}$ that satisfies
\begin{enumerate}
    \item there exists $\alpha_1, \alpha_2 \in \mathcal{K}_{\infty}$ such that for any $x \in \mathbb{R}^n$,
    \begin{align*}
        \alpha_1(|x|_A)\leqslant V(x) \leqslant  \alpha_2(|x|_A).
    \end{align*}
    \item there exist  $\alpha_3 \in \mathcal{K}$ and $\chi\in \mathcal{K}_{\infty}$ such that for all $x\in \mathbb{R}^n$ and $u\in \mathbb{R}^m$,
    \begin{align}\label{lyapsetisscond}
        |x|_A\geqslant \chi(|u|)\implies \dot{V}(x)f(x,u)\leqslant -\alpha_3(|x|_A).
    \end{align}
\hspace{-0.8cm}For compact set $A$, an equivalent representation of (\ref{lyapsetisscond}) is:
    \item there exists $\alpha_3, \alpha_4 \in \mathcal{K}_{\infty}$ such that for all  $x\in \mathbb{R}^n$ and $u\in \mathbb{R}^m$,
    \begin{align*}
        \dot{V}(x)f(x,u)\leqslant -\alpha_3(|x|_A)+\alpha_4(|u|).
    \end{align*}
\end{enumerate}
\end{defn}
\begin{prop}[Theorem 3.4 in \cite{skjetne2004report}]\label{setISStheo}
Assume the closed set $A$ is invariant for (\ref{odeiss}). The system (\ref{odeiss}) is ISS with respect to $A$ if it admits a smooth ISS-Lyapunov function with respect to $A$ and is finite escape-time detectable through $|\cdot|_A$.
\end{prop}
\subsection{Set-stability for systems with time delay}
Consider the autonomous TDS with no perturbations,
\begin{equation}\label{fde1}
    \dot{x}(t)=f(t,x_t),\quad x_0=\phi,
\end{equation}
where $f:\mathcal{C}\to \mathbb{R}^n$ is continuous and $\phi: [-\delta,0]\mapsto \R^n$ is the initial condition. The vector $x(t)$ is the solution at time $t$ while $x_t$ is the state of TDS (\ref{fde1}). 
\begin{defn}[Invariant set w.r.t FDE]\label{invariantsetfde}
A set $A \subseteq C$ is said to be
an invariant set (with respect to (\ref{fde1})) if for any $\phi$ in $A$, there is a solution
$x(\cdot)$ of (\ref{fde1}) that is defined on $(-\infty, \infty)$ such that $x_t \in A$, $\forall t \in (-\infty, \infty)$.
\end{defn}
\begin{defn}[Definition 2.6 in \cite{bernfeld2003stability}] \label{fdeuniformstab}
\begin{enumerate}
    \item $A \subset \mathbb{R}^n$ is referred to as a uniform stable set for (\ref{fde1}) if for every $\epsilon>0$, and $t_0\geqslant 0$, there  exists $\delta'=\delta'(\epsilon)>0$ such  that $\rho(A,\phi)<\delta'$ implies $\rho(A,x_t)<\epsilon$, $\forall t\geqslant t_0$, where $\rho(A,\phi)\triangleq\displaystyle{\inf_{k \in A}}|k-\phi|$.
    \item $A \subset \mathbb{R}^n$ is referred to as uniformly attractive set for (\ref{fde1}) if for some $\eta>0$ and any $\epsilon>0$, there exists $T=T(\epsilon)>0$ such that $\rho(A,x_t)<\epsilon$, $\forall t_0\geqslant 0$, $\rho(A,\phi)<\eta$ and  $t\geqslant t_0+T(\epsilon)$.
    \item  If a set $A \subset \mathbb{R}^n$ is both uniformly stable and uniformly attractive for (\ref{fde1}), it is said to be uniformly asymptotically stable for (\ref{fde1}).
\end{enumerate}

\end{defn} 
 We further introduce FDEs with disturbances and some notations related to the same as given in \cite{teel1998connections}.
  Given a function $w:[-\delta,\infty) \to \mathbb{R}^m $ and $t \in [0,\infty)$,  $w_t(\cdot)$ represents a function from $[0, \delta]$ to $\mathbb{R}^m$ defined by $w_t(\tau)=w(t-\tau)$. We analyze FDEs of the form,
\begin{equation}\label{stdfde}
    \dot{x}(t)=f(t,x_t,w_t), \quad x_0=\xi,
\end{equation}
where $x$ takes values in $\mathbb{R}^n$ and the initial data is continuous. $w$ takes values in $\mathbb{R}^m$ and is bounded and piece-wise continuous. The signal $w$ is an exogenous input. We assume that there exists $T_f > 0$ and an unique maximal solution $x(\cdot)$ defined on $[t_0-\delta, t_0 + T_f)$ for each initial data, input, and starting time $t_0> 0$. The norms are defined as follows
\begin{equation*}\label{normstimedelay}
\begin{aligned}
    |x_t|&\triangleq\max_{-\delta \leqslant s \leqslant 0}|x(t+s)|,\\
    \|x_t\|_{t_0}&\triangleq\sup_{t\geqslant t_0}\{\max_{-\delta \leqslant s \leqslant 0}|x(t+s)|\}.
\end{aligned}
\end{equation*}
The definition of norms with regard to other variables is similar as well. Given continuous functions $x:[-\delta,\infty) \to \mathbb{R}^n$ and $V:[-\delta,\infty)\times \mathbb{R}^n \to \mathbb{R}_{\geqslant 0}$, we use $V(t,x(t))\triangleq V(t)$ and define $V_t(\tau) \triangleq V(t-\tau)$ for $t \geqslant 0$ and $\tau \in [0,\delta]$. Additionally, the upper right hand derivative of $V$ along the solution of (\ref{stdfde}), $x(t)$ is defined as  $D^+V(t)=\lim \displaystyle{\sup_{h \to 0^+}}\frac{V(t+h)-V(t)}{h}$.\newline\newline
To conclude this section we present an instrumental lemma and we introduce a final relevant definition.
\begin{prop}
[Lemma 1 in\cite{teel1998connections}] \label{lemmaimp}
Let $\mu \geqslant 0$ and $\alpha \in \mathcal{K}$. If $V(t) \geqslant \xi$ implies $D^{+}V(t) \leqslant -\alpha(V(t))$, then there exists $\beta \in \mathcal{KL}$ (independent of $\mu$) with $\beta(s,0) \geqslant s$, such that $V (t) \leqslant \max\{\beta(V(t_0),t-t_0),\xi\}$.
\end{prop}

A compact, invariant set $A \subset \mathbb{R}^n$ that contains the origin serves as the cornerstone for the entire discussion in the article. The norms w.r.t set $A$ are defined as follows
\begin{align*}
|x_t|_A&\triangleq\inf_{k\in A}\max_{-\delta \leqslant s \leqslant 0}|k-x(t+s)|,\\
    ||x_t||_A&\triangleq\max_{-\delta \leqslant s \leqslant 0}|x(t+s)|_A=\max_{-\delta \leqslant s \leqslant 0}\inf_{k\in A}|k-x(t+s)|,\\
    \|x_t\|_{{t_0},A}&=\sup_{t\geqslant t_0}\{|x_t|_A\}.
\end{align*}
We also have $\|x_t\|_A\leqslant |x_t|_A$ (under the assumption that system has no finite escape time through $|\cdot|_{A}$) \cite{rudin1976principles}. Invariant set w.r.t (\ref{stdfde}) is similar to that in Definition \ref{invariantsetfde} with $w_t=0$ in (\ref{stdfde}).
\newline
The following definition is set-modified version of [Definition 1 in \cite{teel1998connections}]. 
\begin{defn}\label{isswithgainsetnew}
Let $\gamma \in \mathcal{G},\mu \in \mathbb{R}_{\geqslant 0}$, and $\Delta_x,\Delta_w \in \mathbb{R}_{\geqslant 0} \cup \infty$.
A set $A$ (compact, invariant and contains origin) with respect to (\ref{stdfde}) is said to be uniformly ISS with gain $\gamma$ [and offset $\mu$ and restriction $(\Delta_x, \Delta_w)$] if $|x_0|_A <\Delta_x$ and $\|w_t\|_{{t_0}} <\Delta_w$ imply $T_f = \infty$ and that the following properties
hold uniformly in $t_0 > 0$
\begin{enumerate}
    \item for each $\epsilon>0$ there exists $\delta>0$
such that $|x_0|_A \leqslant \delta$ implies $\|x_t\|_{{t_0},A} \leqslant \max\{\epsilon,\gamma(\|w_t\|_{{t_0}}),\mu\}$
and,
\item for each $\epsilon>0$, $\eta_x \in (0,\Delta_x), \eta_w \in (0,\Delta_w)$  there
exists $T > 0$ such that $|x_0|_A \leqslant \eta_x$ and $\|w_t\|_{{t_0}} \leqslant \eta_w$ imply $\|x_t\|_{{(t_0+T)},A} \leqslant \max\{\epsilon, \gamma(\|w_t\|_{{t_0}}),\mu\}$.
\end{enumerate}
\end{defn}
This forms the standard definition of uniform asymptotic stability for the set $A$ of the FDEs for $\mu=0$ and $w(t)\equiv 0$. It can be proved with the help of Definition \ref{fdeuniformstab}.

\section{Main Results}\label{mainres}
\subsection{Set ISS versions of the Razumikhin-type theorem}\label{main:part1}
In this subsection, we state the set ISS theorems and provide their proof. The following is the global set ISS version of the Razumikhin-type theorem. \cite{efimov2016robustness} presents similar theorem for practical ISS of decomposable sets.
\begin{theo}\label{lyapunovversionset}
If there exist $\alpha_1,\alpha_2 \in \mathcal{K}_{\infty}$, a continuous function $V:[-\delta,\infty)\times\mathbb{R}^n \to \mathbb{R}_{\geqslant 0},\gamma_v,\gamma_w \in \mathcal{G}$ and $\alpha_3 \in \mathcal{K}$ such that 
\begin{enumerate}
    \item $\alpha_1(|x(t)|_A)\leqslant V(t) \leqslant \alpha_2(|x(t)|_A)$,
    \item $V(t) \geqslant \max\{\gamma_v(|V_t|),\gamma_w(|w_t|)\}$\\ $\implies$ $D^{+}V(t) \leqslant -\alpha_3(|x(t)|_A)$,
    \item $\gamma_v(s)<s$ for $s>0$,
\end{enumerate}
then the set $A$ for (\ref{stdfde}) is uniformly globally ISS with gain $\alpha_1^{-1}\circ\gamma_w$.
\end{theo}
\begin{proof}
The pertinent bounding inequalities (when
all signals are bounded) are
\begin{align}\label{firstnew1}
    |V_t| &\leqslant \max \{|V_0|.\phi(t-t_0),\|V\|_{t_0}\},\\\label{secondnew1}
    V(t) &\leqslant \max\{\beta(V(t_0),t-t_0),\gamma_v(||V_t\|_{t_0}),\gamma_w(\|w_t\|_{t_0})\},
\end{align}
where $\phi(s)=0.5(1-sgn(s-\delta))$ and signal $w$ is an exogenous input. It is common to use inequality (\ref{firstnew1}) to constrain the value of $|V_0|$ (e.g., see for example \cite{teel1996nonlinear} or \cite{teel1998connections}). Inequality (\ref{secondnew1}) follows from  Point 2) of the theorem and Proposition \ref{lemmaimp} with $\alpha=\alpha_3 \circ \alpha_2^{-1}$.
Take the sup norm on both the sides of (\ref{firstnew1}) and (\ref{secondnew1}) to obtain
\begin{align}\label{firstnewlyap}
  \|V_t\|_{t_0} &\leqslant \sup_{t \geqslant t_0} \{ \max \{|V_0|.\phi(t-t_0),\|V\|_{t_0}\}\},\\ \nonumber\label{secondnewlyap}
    \|V\|_{t_0} &\leqslant \sup_{t \geqslant t_0} \{ \max\{\beta(\alpha_2(|x_0|_A),t-t_0),\\
    &\qquad\quad\quad \gamma_v(||V_t\|_{t_0}),\gamma_w(\|w_t\|_{t_0})\}\}.
\end{align}
Next substituting (\ref{secondnewlyap}) in (\ref{firstnewlyap}), we get
\begin{align*}
\|V_t\|_{t_0} & \leqslant \max \{\beta(\alpha_2(|x_0|_A),0),\gamma_v(||V_t\|_{t_0}),\\
&\qquad \qquad \qquad \qquad \gamma_w(\|w_t\|_{t_0})\}.
\end{align*}
Following Point 3) of the theorem where $\gamma_v(s)<s$ for $s>0$, we can deduce that, for any $a, b, c \geqslant 0$, if $a \leqslant \max\{b,\gamma_v(a), c\}$, then $a \leqslant \max\{b, c\}$. We now utilize the theorem's Point 1) to establish
\begin{align}\nonumber
    \|V_t\|_{t_0} &\leqslant \max \{\beta(\alpha_2(|x_0|_A),0)),\gamma_w(\|w_t\|_{t_0})\} \\ \nonumber \label{secondfinalnew}
    \implies  \|x_t\|_{{t_0},M} &\leqslant \max \{\alpha_1^{-1} \circ \beta(\alpha_2(|x_0|_A),0)),\\
    &\qquad \qquad\qquad \alpha_1^{-1} \circ\gamma_w(\|w_t\|_{t_0})\}.
\end{align}
Using Definition \ref{isswithgainsetnew}, $T_f=\infty$ and if $|x_0|_A\leqslant \delta'$, for every $\epsilon>0, \exists $ a $\delta'>0$ such that $\beta(\alpha_2(\delta'),0) \leqslant \alpha_1(\epsilon)$. So,  $\|x_t\|_{{t_0},M} \leqslant \max \{\epsilon, \alpha_1^{-1} \circ\gamma_w(\|w_t\|_{t_0})\}$ holds from (\ref{secondfinalnew}).\\\\
For uniform convergence, given strictly positive real numbers $\epsilon,\eta_x,\eta_w$, let $\kappa=\max\{\beta(\alpha_2(\eta_x),0),\gamma_w(\eta_w)\}$, and with $|x_0|_A \leqslant \eta_x$ and $\|w_t\|_{t_0} \leqslant \eta_w$,
we have $ \|V_t\|_{t_0} \leqslant \kappa$.
Let $\rho_1>\delta$ and $\rho_2>0$ be such that $\beta(\kappa,\rho_2) \leqslant \alpha_1(\epsilon)$. Now, using (\ref{firstnew1}) and (\ref{secondnew1}), we have
\begin{align*}\nonumber
    \|V_t\|_{t_0+\rho_1+\rho_2}& \leqslant \|V\|_{t_0+\rho_2}\\
    &\leqslant \max\{\alpha_1(\epsilon), \gamma_v(||V_t\|_{t_0}),\gamma_w(\|w_t\|_{t_0})\}.
\end{align*}
We know, \begin{align*}
  \|V_t\|_{t_0+\rho_1+\rho_2} &\leqslant \displaystyle{\sup_{t \geqslant t_0+\rho_1+\rho_2} }\{ \max \{|V_0|.\phi(t-t_0),\\
  &\qquad \qquad \qquad \qquad \|V\|_{t_0}\}.
\end{align*}
If $t\geqslant t_0+\rho_1+\rho_2$, then $\phi(t-t_0)=0$ which results in
\begin{align*}
     \|V_t\|_{t_0+\rho_1+\rho_2}  \leqslant \|V\|_{t_0+\rho_1+\rho_2}\leqslant \|V\|_{t_0+\rho_2},
\end{align*}
where,
\begin{align*}
    \|V\|_{t_0+\rho_2} &\leqslant \sup_{t\geqslant t_0+\rho_2}\{\max\{\beta(\alpha_2(|x_0|_A),t-t_0),\\
    &\qquad\qquad\quad \gamma_v(||V_t\|_{t_0}),\gamma_w(\|w_t\|_{t_0})\}\}.
\end{align*}
Moreover $\alpha_2(|x_0|_A)\leqslant \beta(\alpha_2(|x_0|_A),0) \leqslant \kappa$. Therefore, $\beta(\alpha_2(|x_0|_A),\rho_2) \leqslant \beta(\kappa,\rho_2) \leqslant \alpha_1(\epsilon)$.\\\\
Since $\gamma_v(s)<s$ for all $s>0$, there exists $n(\kappa,\epsilon)$ such that $\gamma_v^n(\kappa)\leqslant \max\{\alpha_1(\epsilon),\gamma_w(\|w_t\|_{t_0})\}$. We then get to the conclusion that,
\begin{align*}
    \|V_t\|_{t_0+n(\rho_1+\rho_2)}&\leqslant \max\{\alpha_1(\epsilon),\gamma_w(\|w_t\|_{t_0})\}\\
    \implies  \|x_t\|_{{t_0+n(\rho_1+\rho_2)},A}&\leqslant \max\{\epsilon,\alpha_1^{-1}\circ \gamma_w(\|w_t\|_{t_0})\}.
\end{align*}
$\gamma_v^n(\kappa)$ represents composition of $\gamma_v$ with itself for n times. Thus, the second condition of Definition \ref{isswithgainsetnew} also holds, and the set $A$ for (\ref{stdfde}) is uniformly globally ISS with gain $\alpha_1^{-1}\circ\gamma_w$.
\end{proof}
The next version follows from a small-gain argument working with $\|x_t\|_A$ rather than $\|V_t\|_A$. More significantly, it addresses the case where the small-gain condition does not hold on all of $(0,\infty)$. To explain, this situation will arise in our upcoming result on robustness of input-to-state stabilizablity of sets to small time delays at input. 

\begin{theo}\label{isssetRazumikhintype}
Suppose there exists $\alpha_1,\alpha_2 \in \mathcal{K}_{\infty}$, a continuous function $V:[-\delta,\infty)\times\mathbb{R}^n  \to \mathbb{R}_{\geqslant 0},\gamma_v,\gamma_w \in \mathcal{G}$ and $\alpha_3 \in \mathcal{K}$, and non negative real numbers $\mu <\Delta$ such that
\begin{enumerate}
    \item $\alpha_1(|x(t)|_A)\leqslant V(t) \leqslant \alpha_2(|x(t)|_A)$,
    \item $|x(t)|_A \geqslant \max\{\gamma_x(|x_t|_A),\gamma_w(|w_t|)\}$ $\implies$ $D^{+}V(t) \leqslant -\alpha_3(|x(t)|_A)$,
    \item $\alpha_1^{-1} \circ \alpha_2 \circ \gamma_x(s)<s$ for $\mu<s<\Delta$.
\end{enumerate}
Let $\beta \in \mathcal{KL}$ be as in conclusion of Proposition \ref{lemmaimp}
when $\alpha=\alpha_3 \circ \alpha_2^{-1}$. Then the set $A$ is uniformly ISS with gain $\tilde{\gamma}_w\triangleq\alpha_1^{-1}\circ\alpha_2\circ\gamma_w$ (offset $\mu$ and restriction $(\Delta_x,\Delta_w))$ such that $\max\{\alpha_1^{-1}(\beta(\alpha_2(s_1),0)),\tilde{\gamma}_w(s_2)\}<\Delta$ when $s_1 < \Delta_x$, $s_2 < \Delta_w$.
\end{theo}
\begin{proof}
In this case, the pertinent inequalities are
\begin{align}\label{firstxsetnew}
    |x_t|_A &\leqslant \max \{|x_0|_A.\phi(t-t_0),\|x\|_{{t_0},A}\},\\ \nonumber
    |x(t)|_A &\leqslant \max\{\tilde{\beta}(|x(t_0)|_A,t-t_0),\tilde{\gamma}_x(||x_t\|_{{t_0},A}),\\ \label{secondxsetnew}
    &\qquad \quad \quad\tilde{\gamma}_w(\|w_t\|_{{t_0}})\},
\end{align}
where, again $\phi(s)=0.5(1-sgn(s-\delta))$, $\tilde{\beta}(s,t)=\alpha_1^{-1}(\beta(\alpha_2(s),t)$, $\tilde{\gamma}_x=\alpha_1^{-1}\circ\alpha_2\circ\gamma_x$, and $\tilde{\gamma}_w=\alpha_1^{-1}\circ\alpha_2\circ\gamma_w$.
Inequality (\ref{secondxsetnew}) follows by modifying Point 2) of the theorem, i.e., if $V \geqslant \alpha_2(\xi)$ implies $|x|_A \geqslant \xi$. Thus, in accordance with the theorem's Point 2), we will obtain, $V(t)\geqslant \max\{\alpha_2\circ\gamma_x(|x_t|_A),\alpha_2\circ\gamma_w(|w_t|)\}$ $\implies$ $D^{+}V(t) \leqslant -\alpha_3\circ\alpha_2^{-1}(|V(t)|)$. Using Proposition \ref{lemmaimp} and Point 1) of the theorem, we have 
\begin{align*}
V(t) &\leqslant \max\{\beta(V(t_0),t-t_0),\alpha_2\circ\gamma_x(|x_t|_A),\\
&\qquad \qquad \alpha_2\circ\gamma_w(|w_t|)\}\\
\implies  |x(t)|_A 
& \leqslant \max\{\tilde{\beta}(|x_0|_A,0),\tilde{\gamma}_x(||x_t\|_{{t_0},A}),\\
&\qquad \qquad \tilde{\gamma}_w(\|w_t\|_{{t_0}})\}.
\end{align*}
We here make use of the fact that $|x(t_0)|_A\leqslant \|x_0\|_A\leqslant|x_0|_A$.
Using $\max\{\alpha_1^{-1}(\beta(\alpha_2(s_1),0)),\tilde{\gamma}_w(s_2)\}<\Delta$ when $s_1 < \Delta_x$, $s_2 < \Delta_w$, guarantees $|x_0|_A<\Delta$. We have $\alpha_1^{-1}(\beta(\alpha_2(s_1),0)) < \Delta$ when $s_1 < \Delta_x$. For $s_1=|x_0|_A$, we have $\alpha_1^{-1}(\beta(\alpha_2(|x_0|_A),0)) < \Delta$ and,
\begin{align*}
    \alpha_2(|x_0|_A)&\leqslant \beta(\alpha_2(|x_0|_A),0),\\
\alpha_1(|x_0|_A)& \leqslant\alpha_2(|x_0|_A)\leqslant \beta(\alpha_2(|x_0|_A),0).
\end{align*}
The existence of the aforementioned truncation guarantees $|x_t|_A<\Delta$, and $\|x_t\|_{{t_0},A}<\Delta$.
Taking sup norm on both sides of (\ref{firstxsetnew}) and (\ref{secondxsetnew}) and substituting the bound of $\|x\|_{{t_0},A}$ in (\ref{secondxsetnew}), we obtain
\begin{align}\label{secondfinalXnew}
  \|x_t\|_{{t_0},A}  &\leqslant \max \{\tilde{\beta}((|x(t_0)|_A),0),\mu,\tilde{\gamma}_w(\|w_t\|_{{t_0}})\}.
\end{align}
 Following Point 3) of the theorem where $\tilde{\gamma}_x(s)<s$ for $s>0$, we can deduce that for $a, b, c\in [0, \Delta)$, $a \leqslant \max\{b,\tilde{\gamma}_x(a),c\}$
imply $a\leqslant \max\{b,\mu, c\}$. Moreover, we are aware that $\tilde{\gamma}_w (s_2) <\Delta$.
Thus, the first condition of the Definition \ref{isswithgainsetnew} holds if $|x_0|_A\leqslant k$, for every $\epsilon>0$, $\exists$ a $k>0$ such that $\tilde{\beta}(k,0) \leqslant (\epsilon)$. From (\ref{secondfinalXnew}) we have, $\|x_t\|_{{t_0},A}  \leqslant \max \{\tilde{\beta}((|x_0|_A),0),\mu,\tilde{\gamma}_w(\|w_t\|_{{t_0}})\}$ and, $\|x_t\|_{{t_0},A} \leqslant \max \{\epsilon, \tilde{\gamma}_w(\|w_t\|_{{t_0}}), \mu\}$.\\\\
In order to demonstrate uniform convergence, we follow a similar procedure to that used in Theorem \ref{lyapunovversionset}. For given strictly positive real values $\epsilon,\eta_x,\eta_w$, let $\kappa=\max\{\tilde{\beta}(\eta_x,0),\tilde{\gamma}_w(\eta_w),\mu\}$, where $|x_0|_A \leqslant \eta_x$ and $\|w_t\|_{{t_0}} \leqslant \eta_w$ to obtain $\|x_t\|_{{t_0},A} \leqslant \kappa$.
Let $\rho_2>0$ be such that $\tilde{\beta}(\kappa,\rho_2) \leqslant (\epsilon)$. Also let $\rho_1>\mu$. Using (\ref{firstxsetnew}) and (\ref{secondxsetnew}), we can then generate
\begin{align*}\nonumber
    \|x_t\|_{{(t_0+\rho_1+\rho_2)},A}& \leqslant \|x\|_{{(t_0+\rho_2)},A}\\
    &\leqslant \max\{\epsilon, \tilde{\gamma}_x(||x_t\|_{{t_0},A}),\tilde{\gamma}_w(\|w_t\|_{{t_0}})\}.
\end{align*}
Since $\tilde{\gamma}_x(s)<s$ for all $\mu<s<\Delta$, there exists $n(\kappa,\epsilon)$ such that $\tilde{\gamma}_x^n(\kappa)\leqslant \max\{\epsilon,\tilde{\gamma}_w(\|w_t\|_{{t_0}}),\mu\}$. We may therefore deduce that
\begin{align*}
    \|x_t\|_{{(t_0+n(\rho_1+\rho_2))},A}&\leqslant \max\{\epsilon,\tilde{\gamma}_w(\|w_t\|_{{t_0}}),\mu\}.
\end{align*}
Thereby, we demonstrate that the second condition of the Definition \ref{isswithgainsetnew} also holds here and set $A$ for (\ref{stdfde}) is uniformly ISS with gain $\tilde{\gamma}_w$[offset $\mu$ and restriction $(\Delta_x,\Delta_w)$].
\end{proof}
\subsection{Set input-to-state stabilizibility}\label{main:part2}
Next we state a lemma pertinent to this subsection.
\begin{lemma}\label{lemmaforB(x)}
Consider a system 
    \begin{align} \label{doubleinput2}
        \dot{x}=\phi(x,u_1,u_2),
    \end{align}
     where, $x\in \mathbb{R}^n$ denotes the vector of state variables, $u_1 \in \mathbb{R}^s,u_2 \in \mathbb{R}^m$ denote vectors of input variables (disturbances), and $\phi$ is
locally Lipschitz on $\mathbb{R}^n\times \mathbb{R}^s\times\mathbb{R}^m$.\\
    If the system (\ref{doubleinput2}) with $u_2\equiv0$ is ISS w.r.t set $A$ with gain $\gamma_{u_1}$,
then there exists a $m \times  m$ matrix $B(x,u_1)$ of smooth functions, invertible for all $x \in \mathbb{R}^n$, $u_1 \in \mathbb{R}^s$ that satisfies
$B(x,u_1) \equiv I_{m\times m}$ in a neighborhood of the origin of set $A$,
and a function $\gamma_{u_2}\in \mathcal{K}_{\infty}$, such that the system 
\begin{align}\label{eqndoubleinput}
    \dot{x}=\phi(x,u_1,B(x,u_1)u_2),
\end{align}
is ISS w.r.t set $A$ with gain $(\gamma_{u_1},\gamma_{u_2})$. More specifically, $\exists$ $\alpha_i \in \mathcal{K}$ for $i=1,2,3$, such that $\alpha_1(|x|_A) \leqslant V(x) \leqslant \alpha_2(|x|_A)$ and the following hold true
\begin{align*}
    |x|_A &\geqslant \max\{\tilde{\gamma}_{u_1}(|u_1|),\tilde{\gamma}_{u_2}(|u_2|)\}\\ &\implies \frac{\partial V}{\partial x} \phi(x,u_1,B(x,u_1)u_2)\leqslant -0.5\alpha_3(|x|_A),
\end{align*}
where, $\tilde{\gamma}_{u_1}(s)=\alpha_2^{-1}(\alpha_1(\gamma_{u_1}(s)))$ and $\tilde{\gamma}_{u_2}(s)=\alpha_2^{-1}(\alpha_1(\gamma_{u_2}(s)))$.
\end{lemma}
\begin{proof}
    We proceed along the lines of the proof of Lemma 1 in \cite{christofides1996singular}, that there exist a smooth function $V: \mathbb{R}^n \to \mathbb{R}_{\geqslant 0}$, $\alpha_1, \alpha_2, \alpha_3 \in \mathcal{K}_{\infty}$, and $\tilde{\gamma}_{u_1} \in \mathcal{K}$ such that $\alpha_1(|x|_A) \leqslant V(x) \leqslant \alpha_2(|x|_A)$ holds and 
\begin{align*}
    |x|_A \geqslant \tilde{\gamma}_{u_1}(|u_1|) \implies \dot{V}=\frac{\partial V}{\partial x} \phi(x,u_1,0)\leqslant -\alpha_3(|x|_A),
\end{align*}
where $\gamma_{u_1}(s)=\alpha_1^{-1}\circ \alpha_2 \circ \tilde{\gamma}_{u_1}(s)\in \mathcal{K}$. We define $\tilde{u}_2=B(x,u_1)u_2$ and calculate the time derivative of the function $V$ along the trajectories of (\ref{eqndoubleinput}),
\begin{align}\nonumber
    \dot{V}&=\frac{\partial V}{\partial x} \phi(x,u_1,B(x,u_1)u_2)\\ \label{vdotinlemma2}
    &=\frac{\partial V}{\partial x} \phi(x,u_1,0)+\frac{\partial V}{\partial x} (\phi(x,u_1,\tilde{u}_2)-\phi(x,u_1,0)).
\end{align}
As $\phi$ is locally Lipschitz, there exist a $\psi \in \mathcal{K}_{\infty}$ and a positive real number $L\geqslant 1$ such that using (\ref{vdotinlemma2}), we get
\begin{align*}
    |x|_A &\geqslant \tilde{\gamma}_{u_1}(|u_1|)\\
    \implies \dot{V}&\leqslant -\alpha_3(|x|_A)+|\tilde{u}_2|(L+\psi(\max\{|x|,|u_1|,|\tilde{u}_2|\})).
\end{align*}
Let us now define $\tilde{\gamma}_{u_2}(s)\triangleq \max\{s,\alpha_3^{-1}(2(1+L)(s))\} \in \mathcal{K}_{\infty}$ and consider $b(s)$ to be a smooth function which satisfies $b(s)=1$ in neighborhood of origin of set $A$ (i.e., $\exists$ a $\delta_1>0$ such that $b(s)=1$, $\forall s \in [0,\delta_1])$ and specifically is chosen so that the following inequality 
\begin{align}\label{bfcn}
    0< b(s) \leqslant \min\Big\{\frac{1}{0.5+\psi(s)},1\Big\}
\end{align}
holds for all $s \in \mathbb{R}_{\geqslant 0}$. For $X=[x^{\trp}, u_1^{\trp}]^{\trp}$, we assert that the matrix  $B(x,u_1)\triangleq b(|X|)I_{m\times m}$ and the function $\gamma_{u_2}(s)=\alpha_1^{-1}\circ \alpha_2 \circ \tilde{\gamma}_{u_2}(s)$ satisfies the requirement of the lemma, i.e., 
\begin{equation}\label{maincondnn}
\begin{aligned}
      |x|_A & \geqslant \max\{\tilde{\gamma}_{u_1}(|u_1|),\tilde{\gamma}_{u_2}(|u_2|)\}\\  \implies& \frac{\partial V}{\partial x} \phi(x,u_1,B(x,u_1)u_2)\leqslant -0.5\alpha_3(|x|_A).
\end{aligned}
\end{equation}
Using the fact from (\ref{bfcn}), that $b(|X|) \leqslant 1$, we have
\begin{align*}
  |x|_A & \geqslant  \tilde{\gamma}_{u_1}(|u_1|)\\
    \implies \dot{V}\leqslant& -\alpha_3(|x|_A)\\
    &+b(|X|)|u_2|(L+\psi(\max\{|x|,|u_1|,|u_2|\})).
\end{align*}
Now,
\begin{align*}
    |x|_A \geqslant & \max\{\tilde{\gamma}_{u_1}(|u_1|),|u_2|\}\\ 
     \implies \dot{V}\leqslant & -\alpha_3(|x|_A)+b(|X|)|u_2|(L+\psi(\max\{|x|,|u_1|\})).
\end{align*}
We use the fact from \cite{chaillet2006stability} that $|x|_A \leqslant |x|$ when a compact set contains the origin, which implies $|u_2|\leqslant |x|_A\leqslant |x|$ for the above inequality. So,
\begin{align*}
     |x|_A \geqslant& \max\{\tilde{\gamma}_{u_1}(|u_1|),|u_2|\}\\ 
     \implies \dot{V}\leqslant& -\alpha_3(|x|_A)+b(|X|)|u_2|(L+\psi(|X|)).
\end{align*}
From (\ref{bfcn}) we have, $b(s)\leqslant 1$ and $b(s)\psi(s)\leqslant 1$ for $s\geqslant 0$ and so,
\begin{align*}
    |x|_A \geqslant & \max\{\tilde{\gamma}_{u_1}(|u_1|),|u_2|\}\\ 
     \implies & \dot{V}\leqslant -0.5\alpha_3(|x|_A) -0.5\alpha_3(|x|_A) +(L+1)|u_2|.
\end{align*}
which can be rewritten as
\begin{align*}
    |x|_A \geqslant & \max\{\tilde{\gamma}_{u_1}(|u_1|),|u_2|,\alpha_3^{-1}(2(L+1)(|u_2|))\}\\ 
     \implies & \dot{V}\leqslant -0.5\alpha_3(|x|_A).
\end{align*}
where $\tilde{\gamma}_{u_2}(s)\triangleq \max\{s,\alpha_3^{-1}(2(1+L)(s))\}$ and $\gamma_{u_2}(s)=\alpha_1^{-1}\circ \alpha_2 \circ \tilde{\gamma}_{u_2}(s)$. Hence, (\ref{maincondnn}) is satisfied.
\end{proof}
\subsection{Robustness of set ISS to small time delays at input}\label{main:part3}
The key result of this article is now ready to be stated.
We first give the results of the robustness analysis as provided in \cite{efimov2016robustness}, but without the use of any practical constants. Additionally, we will compare the outcomes for the robustness of set ISS to small input time delays.\newline\newline
Consider the following system
\begin{equation}\label{sysefimov}
    \dot{x}(t)=f(x(t),u(t)),
\end{equation}
where $\tilde{\delta} >0$, $x\in \mathbb{R}^n$, and $u\in \mathbb{R}^m$. 
\begin{theo}\label{robustness2}
If set $A$ for (\ref{sysefimov}) is input-to-state stabilizable with some gain $\tilde{\gamma}$ in the absence of delay, then there exists some $\tilde{\delta} >0$ such that the set $A$ for $\dot{x}(t)=f(x(t ),u+ g(x_t))$ (where $g$ is a continuous function) is uniformly ISS with gain $\tilde{\gamma}'$.
\end{theo}
\begin{proof}
If (\ref{sysefimov}) is ISS w.r.t. set $A$, then it satisfies the following requirements,
    \begin{align}\label{firstineq}
    \alpha_1(|x|_A)&\leqslant V(x) \leqslant \alpha_2(|x|_A),\\ \label{secineq}
    \text{and } DV(x)f(x,u)&\leqslant -\alpha_3(|x|_A)+\gamma(|u|),
    \end{align}
    where the gain $\tilde{\gamma}\triangleq\alpha_1^{-1}\circ \alpha_2\circ\gamma_1$, $\gamma_1=\alpha_3^{-1}(\frac{1}{1-\epsilon}\gamma)$ and $\epsilon \in (0,1)$. Using (\ref{firstineq}), we know
    \begin{align}\label{replace}
    -\alpha_3(|x|_A)&\leqslant -\alpha_3 (0.5\alpha_2^{-1}(V(x)).
    \end{align}
    Using (\ref{replace}) in (\ref{secineq}), we obtain
    \begin{align} \label{finallyapefimov}
    DV(x)f(x,u)&\leqslant -\alpha_4(V(x)) +\gamma(|u|),
\end{align}
where $\alpha_4(s)=\alpha_3( 0.5\alpha_2^{-1}(s))$. Now that the system is time delayed, assume the input $u$ has two terms $u_1$ and $u_2$, and $u_2$ is a function of $x_t$ for some $\delta>0$, i.e.,
\begin{align}\label{delay2}
    u=u_1+u_2, u_2=g(x_t),
\end{align}
where $g$ is a continuous function and
\begin{align*}
   |g(x_t)|\leqslant k(|V_t|)  
\end{align*}
for some $k \in \mathcal{G}$. For ease of notation, additionally denote $u=u_1$, then the system (\ref{sysefimov}) is transformed to 
\begin{align*}
    \dot{x}(t)=f(x(t),u+g(x_t)),
\end{align*}
and we replace (\ref{delay2}) in (\ref{finallyapefimov}) to get
\begin{align*}\label{firstresineq}
     D^+V(t)&\leqslant -\alpha_4(V(t)) +\gamma(2|u_t|)+\gamma(2k(|V_t|)),
\end{align*}
which can be rewritten as
\begin{align*}
    V(t)&\geqslant \max\{\gamma_v(|V_t|),\gamma_u(|u_t|)\}\\
    \implies D^+V(t) &\leqslant -0.5\alpha_5(|x(t)|_A),
\end{align*}
where $\gamma_v(s)=\alpha_4^{-1}(2\gamma(4k(s))$, $\gamma_u(s)=\alpha_4^{-1}(2\gamma(4(s))$, and $\alpha_5(s)=\alpha_4(\alpha_2(s))$. The three requirements of Theorem \ref{lyapunovversionset} are satisfied if $\gamma_v(s)<s$ for all $s>0$, with ISS gain $\tilde{\gamma}'=\alpha_1^{-1}\circ \gamma_u$, which concludes the proof.
\end{proof}
We will now utilize Theorem \ref{isssetRazumikhintype}
to demonstrate the relaxed version of robustness of set ISS before comparing the two results. \newline\newline
Consider the system
\begin{equation}\label{isssysnew}
    \dot{x}(t)=f(x(t),u(t-\tilde{\delta}),w(t)),
\end{equation}
where $\tilde{\delta} >0$, $x\in \mathbb{R}^n$, $u \in \mathbb{R}^c$ and $w\in \mathbb{R}^m$. If there exists $\alpha_1,\alpha_2,\alpha_3 \in \mathcal{K}_{\infty}$, smooth functions $k:\mathbb{R}^n \to \mathbb{R}^m$, $V:\mathbb{R}^n \to \mathbb{R}_{\geqslant 0}$ and $\gamma \in \mathcal{G}$ such that $\alpha_1(|x(t)|_A)\leqslant V(t) \leqslant \alpha_2(|x(t)|_A)$, $|x|_A\geqslant \gamma(|w|)$ $\implies$ $\frac{\partial V}{\partial x}f(x,k(x),w) \leqslant -\alpha_3(|x(t)|_A)$, we say that set $A$ for (\ref{isssysnew}) is input-to-state stabilizable with gain $\tilde{\gamma}\triangleq \alpha_1^{-1}\circ \alpha_2\circ\gamma$ in the absence of delay \cite{gasparri2014set}.
\begin{theo}\label{robustness1}
If set $A$ for (\ref{isssysnew}) is input-to-state stabilizable with gain $\tilde{\gamma}$ in the absence of delay, then there exists $\delta^*>0$ such that for all $\tilde{\delta} \in(0,\delta^*)$, the set $A$ is uniformly ISS with gain $\tilde{\gamma}$, nonzero offset and nonzero restriction. In addition, the offset becomes arbitrarily small and the restriction becomes arbitrarily large as $\tilde{\delta}$ approaches zero.
\end{theo}
\begin{proof}
Lemma \ref{lemmaforB(x)} states that if the system (\ref{isssysnew}) is ISS w.r.t. a set $A$ in absence of delay, then there exists a smooth invertible $m \times m$ matrix $G$ and $\gamma_{\theta} \in \mathcal{K}_{\infty}$ such that the system 
\begin{align}\label{xdotgnew}
    \dot{x}=f(x,k(x)+G(x,w)\theta,w),
\end{align}
satisfies
\begin{align}\nonumber\label{condnmainnew}
    |x(t)|_A &\geqslant \max\{\gamma_{\theta}(|\theta(t)|),\gamma(|w(t)|)\} \\
    \implies \dot{V} &\leqslant -0.5\alpha_3(|x(t)|_A).
\end{align}
The original system (\ref{isssysnew}) can be expressed as (\ref{xdotgnew}) with $\theta(t)=G(x(t),w(t))^{-1}[k(x(t-\tilde{\delta}))-k(x(t))]$. The function $k$ and $f$ are locally Lipschitz and $\xi(t)= k(x(t-\tilde{\delta}))-k(x(t))$. Since $k$ is locally Lipschitz, we have,
\begin{align*}
    |\xi(t)| &\leqslant L|x(t-\tilde{\delta})-x(t)|\leqslant L\Big|\int_{t-\tilde{\delta}}^{t}\frac{dx}{ds}ds\Big|\\
    &\leqslant L\Big|\int_{t-\tilde{\delta}}^{t}f(x(s),x(s-\tilde{\delta}),w(s))ds\Big|,
\end{align*}
for some Lipschitz constant $L>0$.  We arrive at the last inequality using (\ref{isssysnew}). When we apply the integral mean value theorem to the equation above, we get, 
\begin{align*}
     |\xi(t)| &\leqslant L \tilde{\delta} |f(x(c),x(c-\tilde{\delta}),w(c))|,
\end{align*}
where, $c \in (t-\tilde{\delta},t)$.
Therefore, a $\gamma_1^* \in \mathcal{K}$ will exist such that
$|\xi(t)| \leqslant  \tilde{\delta} \displaystyle\max\{\gamma_1^*(|x_t|_A),\gamma_1^*(|w_t|_A)\}$. One of such constructions of $\gamma_1^* \in \mathcal{K}$ can be as follows,
\begin{align*}
\gamma_1^*(r_1)&=\max_{|\kappa|_A \leqslant r_1,|\lambda|_A \leqslant r_1,|\psi|_A \leqslant r_1}|f(\kappa,\lambda,\psi)|,
\end{align*}
where we assume that $f(\kappa,\lambda,\psi)=0$ when $\kappa, \lambda, \psi \in A$ (this ensures $\gamma_1^*(0)=0$). We will have,
\begin{align*}
    &|f(x(c),x(c-\tilde{\delta}),w(c))| \\
    &\leqslant \max_{|\kappa|_A \leqslant |x(c)|_A,|\lambda|_A \leqslant |x(c-\tilde{\delta})|_A,|\psi|_A \leqslant |w(c)|_A}|f(\kappa,\lambda,\psi)|\\
     &\leqslant  \max_{|\kappa|_A \leqslant |x_t|_A,|\lambda|_A \leqslant |x_t|_A,|\psi|_A \leqslant |w_t|_A}|f(\kappa,\lambda,\psi)|\\
   &\leqslant  \max\Big\{\max_{|\kappa|_A \leqslant |x_t|_A,|\lambda|_A \leqslant |x_t|_A,|\psi|_A\leqslant |x_t|_A}|f(\kappa,\lambda,\psi)|,\\
    &\qquad \quad\max_{|k|_A \leqslant |w_t|_A,|\lambda|_A \leqslant |w_t|_A,|\psi|_A \leqslant |w_t|_A}|f(\kappa,\lambda,\psi)|\Big\}\\
    &\leqslant \max\{\gamma_1^*(|x_t|_A),\gamma_1^*(|w_t|_A)\},
\end{align*}
where, $|x_t|_A=\displaystyle{\inf_{k\in A}\max_{-2\tilde{\delta} \leqslant s \leqslant 0}}(|k-x(t+s)|)$, and $|w_t|_A=\displaystyle{\inf_{k\in A}\max_{-2\tilde{\delta} \leqslant s \leqslant 0}}(|k-w(t+s)|)$. The above calculation shows
\begin{align*}
    |\xi(t)|&\leqslant  L \tilde{\delta} |f(x(c),x(c-\tilde{\delta}),w(c))|\\
    &\leqslant L\tilde{\delta}\max\{\gamma_1^*(|x_t|_A),\gamma_1^*(|w_t|_A)\}.
\end{align*}
Additionally, some $\gamma_1, \gamma_2\in \mathcal{K}$ will exist such that 
\begin{align}\label{eqnthetat}
    |\theta(t)| \leqslant \tilde{\delta}  \max\{\gamma_1(|x_t|_A),\gamma_2(|w_t|_A)\}.
\end{align}
 $A$ is a compact set which contains origin so, $|w_t|_A \leqslant |w_t|$, thus (\ref{eqnthetat}) can be written as
\begin{equation*}
     |\theta(t)| \leqslant \tilde{\delta}  \max\{\gamma_1(|x_t|_A),\gamma_2(|w_t|)\}.
\end{equation*}
Defining $\hat{\gamma} \in \mathcal{G}$ by $\hat{\gamma}\triangleq\max\{\gamma_{\theta}(\tilde{\delta}.\gamma_2(s)),\gamma(s)\}$, it follows from (\ref{condnmainnew}) that
\begin{align*}
    |x(t)|_A &\geqslant \max\{\gamma_{\theta}(\tilde{\delta}.\gamma_1(|x_t|_A)),\hat{\gamma}(|w_t|)\}\\
    &\implies \dot{V}\leqslant -0.5 \alpha_3(|x(t)|_A).
\end{align*}
The subsequent analysis follows from \cite{teel1998connections}.
Since $\gamma_{\theta}(0)=0$, for each pair of strictly positive real numbers $\mu$, and $\Delta$, there exists $\delta^* >0$ such that $\tilde{\delta} \in (0,\delta^*)$ implies
\begin{align*}
    \alpha_1^{-1}\circ\alpha_2 \circ \gamma_{\theta}(\tilde{\delta}.\gamma_1(s))<s, \quad \forall s \in (\mu,\Delta).
\end{align*}
Given an arbitrarily small offset $\mu>0$ and an arbitrarily large (finite) restriction $(\Delta_x, \Delta_w)$, it follows from the Theorem \ref{isssetRazumikhintype} that there exists $\delta^* >0$ such that $\tilde{\delta} \in (0,\delta^*)$  indicates the set $A$ is uniformly
ISS with gain $ \alpha_1^{-1}\circ\alpha_2 \circ \hat{\gamma}$. Given the offset $\delta$ and the restriction
$\Delta_w$ it is always feasible to further refine $\tilde{\delta}$ such that
\begin{align*}
     \alpha_1^{-1}\circ\alpha_2 \circ \gamma_{\theta}(\tilde{\delta}.\gamma_2(\Delta_w)) \leqslant \mu,
\end{align*}
with $\mu$ being arbitrarily small.
At this point, $\alpha_1^{-1}\circ\alpha_2 \circ \hat{\gamma}=\max\{\delta,\alpha_1^{-1}\circ\alpha_2 \circ \gamma\}$. Thus, $\alpha_1^{-1}\circ\alpha_2 \circ \hat{\gamma}$ can be replaced by $\alpha_1^{-1}\circ\alpha_2 \circ \gamma=\tilde{\gamma}$.
\end{proof}
\subsection{Comparative Review}\label{main:part4}
We compare our work with \cite{efimov2016robustness} because, to the best of our knowledge, \cite{efimov2016robustness} is the only pertinent study that analyzes the robustness of set ISS to small input delays. The novel contributions of our work w.r.t \cite{efimov2016robustness} are as below
\begin{itemize}
\item The robustness of the set ISS of a nonlinear system to input is demonstrated in the existing literature, in particular \cite{efimov2016robustness}, where the input is dependent on a time delayed state as an additive effect, which may not always be the case in practical applications. We are putting forth a more general setup to address this discrepancy and demonstrate the set ISS of nonlinear system's robustness to input time delays. In mathematical terms, \cite{efimov2015weighted} prove that if $\dot{x}(t)=f(x(t),w(t))$ is set ISS, then $\dot{x}(t)=f(x(t ),w+ g(x_t))$ will also be set ISS for some time delay $\delta>0$. On the other hand, our study establishes that if $\dot{x}(t)=f(x(t),u(t),w(t))$ is set ISS for some $u=k(x(t))$, then the system $\dot{x}(t)=f(x(t),k(x_t),w(t))$ is also set ISS for some small time delay $\delta>0$, with $k,g$ being continuous functions. 
    \item The time delayed state function is yet again constrained in \cite{efimov2016robustness} as $|g(x_t)|\leqslant h(|V_t|)$, for some $h \in \mathcal{K}_{\infty}$, however in our work, the delayed states are not subject to any such limitations. This condition indicated in \cite{efimov2016robustness} is applicable to fewer systems to show set ISS robustness to time delay at the input as opposed to our method. We will demonstrate via examples that it will be difficult for this criterion in \cite{efimov2016robustness} to hold in some scalar systems.
\end{itemize}
The comparisons are based on Section 4.1 in \cite{efimov2016robustness} (which has also been restated in our article as Theorem \ref{robustness2}, with practical constants as zero). Therefore, we essentially compare Section 4.1 of \cite{efimov2016robustness} with respect to Theorem \ref{robustness1} of our paper in theory and with the aid of illustrative examples in the following section.
\section{Comparison using Examples}\label{examplesection}
This section uses examples to show the robust set ISS analysis of a nonlinear system with delays described in Section \ref{main:part3}. First of all, both examples show how some of the presumptions in Theorem \ref{robustness2} might be challenging to meet. Second, we demonstrate the robustness of the set ISS with delay using Theorem \ref{robustness1}.\\\\
The first example demonstrates the robustness of the set ISS to small time delays at the input by using a compact set that just contains the origin.
\subsection{Example 1}\label{exam1}
Consider a nonlinear oscillator:
\begin{equation}\label{nonlinearoscsys}
\begin{aligned}
\left\{\begin{array}{l}
\dot{x}_{1}=x_{2} \\
\dot{x}_{2}=-k\left(x_{1}\right)-\mu\left(x_{2}\right)+w,
\end{array}\right.
\end{aligned}
\end{equation}
where the continuous functions $k: \mathbb{R} \rightarrow \mathbb{R}$ and $\mu: \mathbb{R} \rightarrow \mathbb{R}$ satisfy the following conditions $k(0)=\mu(0)=0$ and that for any $r \neq 0$ we have
$$
r k(r)>0, \quad r \mu(r)>0 .
$$
$w$ stands for a disturbing force. Under the above assumptions the origin is the only one equilibrium point of the unperturbed $(w=0)$ system.
Let $\mu(s)=\mu s$ be linear, where $\mu>0$ and $k$ be continuous with $k(0)=0$ and $r k(r)>0$ for any $r \in \mathbb{R}$ and in addition $\liminf _{r \rightarrow \pm \infty} k(r) \neq$ 0, and $\eta \in \mathcal{K}_{\infty}$ be such that $k(|s|) \geqslant\eta(|s|)$ for any real $s \neq 0$. Consider the following Lyapunov function and it's time derivative:
\begin{align*}
    V(x)&=\frac{\mu^{2}}{2} x_{1}^{2}+\mu x_{1} x_{2}+x_{2}^{2}+2 \int_{0}^{x_{1}} k(r) d r\\
    \dot{V}(x, u)&=-\mu\left(x_{1} k\left(x_{1}\right)+x_{2}^{2}\right)+\left(\mu x_{1}+2 x_{2}\right) w \\
& \leq-\frac{1}{2} \mu\left(\left|x_{1}\right| \eta\left(\left|x_{1}\right|\right)+x_{2}^{2}\right)
\end{align*}
if
\begin{align}\label{condnissexample1new}
\frac{1}{2} \mu\left(\left|x_{1}\right| \eta\left(\left|x_{1}\right|\right)+x_{2}^{2}\right) \geq\left(\mu\left|x_{1}\right|+2\left|x_{2}\right|\right)|w|. 
\end{align}
It is well known from \cite{dashkovskiy2019practical} that
\begin{align}\label{inputcondn}
|x| \geqslant\max \left\{\sqrt{1+\frac{4}{\mu^{2}}} \frac{8|w|}{\mu}, \sqrt{1+\frac{\mu^{2}}{4}} \eta^{-1}\left({4}|w|\right)\right\}\triangleq \gamma(|w|)
\end{align}
implies (\ref{condnissexample1new}). So, we have $V$ to be an ISS-Lyapunov function. By Definition \ref{setisslyap} the system is ISS with the gain $\gamma \in \mathcal{K}_{\infty}$ defined in (\ref{inputcondn}).\newline\newline
We can also represent the nonlinear oscillator system (\ref{nonlinearoscsys}) as
\begin{align*}
    \dot{x}=f(x)+w_1,
\end{align*}
where $x=(x_1,x_2)^{\top}\in \mathbb{R}_2$, $w_1=(0,w)^{\top}\in\mathbb{R}^2$ and $f(x)=(-x_2,-k(x_1)-x_2)^{\top}\in \mathbb{R}^2$. The subsequent equation with an input delay perturbation is
\begin{align*}
     \dot{x}=f(x)+w_1+g(x_t),
\end{align*}
with $g(x_t)=\begin{pmatrix}
 -x_2(t)+x_2(t-\tau)\\x_2(t)-x_2(t-\tau),
\end{pmatrix}$, $\tau>0$, and $\mu=1$, which results in the following time delayed modification of (\ref{nonlinearoscsys}):
\begin{equation*}
\begin{aligned}
\left\{\begin{array}{l}
\dot{x}_{1}(t)=x_{2}(t-\tau) \\
\dot{x}_{2}(t)=-k\left(x_{1}(t)\right)-\mu x_{2}(t-\tau)+w(t).
\end{array}\right.
\end{aligned}
\end{equation*}
Applying the mean value theorem we obtain
\begin{align*}
     |g(x_t)|=\sqrt{2}\tau\max_{t-\tau\leqslant s\leqslant t}|-k\left(x_{1}(\phi)\right)- x_{2}(\phi-\tau)+w(\phi)|,
\end{align*}
where $\phi \in [s-\tau,s]$ and $|x_t|=\sup_{-2\tau \leqslant s \leqslant 0}|x(t+s)|$. It is difficult to establish the requirement that there exists a $h \in\mathcal{K}_{\infty}$, such that $|g(x_t)|\leqslant h(|V_t|)$, even in situations where Theorem \ref{robustness2} is applicable. This is because  we don't have the information of shared relationship between $w(\phi)$ and $x_1(\phi),x_2(\phi-\tau)$. \newline\newline We propose that the ISS of the time delayed system can be demonstrated using results from ISS robustness with small time delays at the input in \cite{teel1998connections} (extended for sets in Theorem \ref{robustness1}) rather than having to go through the tedious process of validating the above requirements of Theorem \ref{robustness2}. In accordance with (\ref{nonlinearoscsys}), we take into consideration the following system
\begin{equation}\label{nonlinearoscourapp}
\begin{aligned}
\left\{\begin{array}{l}
\dot{x}_{1}(t)=u \\
\dot{x}_{2}(t)=-k\left(x_{1}(t)\right)-\mu u+w(t),
\end{array}\right.
\end{aligned}
\end{equation}
with feedback $u=x_2$.
Alternatively, we can write the above nonlinear oscillator  (\ref{nonlinearoscourapp}) with $\mu=1$ as,
\begin{align*}
    \dot{x}=f(x)+u+w_1,
\end{align*}
where $x=(x_1,x_2)^{\top}\in \mathbb{R}_2,u=(u_1,u_2)^{\top}\in\mathbb{R}^2$, $w_1=(0,w)^{\top}\in\mathbb{R}^2$ and $f(x)=(0,-k(x_1))^{\top}\in \mathbb{R}^2$. We choose $u=(u_1,u_2)^{\top}=(x_2,-x_2)^{\top}$, such that the nonlinear oscillator (\ref{nonlinearoscourapp}) is ISS without any time delay at input. If we have time delay introduced in the feedback, we will then have:
\begin{align}\label{exnewourdelaysys}
    \dot{x}=f(x)+\begin{pmatrix}x_2(t-\tau)\\-x_2(t-\tau)\end{pmatrix}+w_1.
\end{align}
Given that the aforementioned system (\ref{nonlinearoscourapp}) is ISS without any time delay, (\ref{exnewourdelaysys}) rewritten as,
\begin{align*}
    \dot{x}=f(x)+u+G(x,w)\theta+w_1,
\end{align*}
where $G(x,w)=I_2$ and 
\begin{align*}
    \theta=(\theta_1,\theta_2)^{\top}=(x_2(t-\tau)-x_2(t),-(x_2(t-\tau)-x_2(t)))^{\top}
\end{align*}
satisfies the following
\begin{align}\label{eqn:exnewcond1}
    |x(t)|_A&=|x(t)|\geqslant \max \{\gamma(|w(t)|),\gamma_{\theta}(|\theta(t)|)\},\\ \nonumber
    \gamma(|w(t)|)&\triangleq \max \left\{\sqrt{1+4} ({8|w|}), \sqrt{1+\frac{1}{4}} \eta^{-1}\left({4}|w|\right)\right\}\\ \nonumber
    \gamma_{\theta}(|\theta(t)|)&\triangleq \max \Big\{ \max\{10(\eta^{-1}(|\theta|))^4, 10(\eta^{-1}(|\theta|))^2\},\\ \nonumber
    &\qquad \qquad \qquad \qquad \qquad 24\sqrt{10}(|\theta|)\Big\}\\ \nonumber
    \implies \dot{V}&\leqslant -\frac{1}{4} \left(\left|x_{1}\right| \eta\left(\left|x_{1}\right|\right)+x_{2}^{2}\right).
\end{align} 
We will prove the claim made above using the same Lyapunov function as we did for nonlinear oscillator (\ref{nonlinearoscsys}):
\begin{align*}
    V(x)&=\frac{1}{2} x_{1}^{2}+x_{1} x_{2}+x_{2}^{2}+2 \int_{0}^{x_{1}} k(r) d r\\
    \dot{V}(x, w)&=-\left(x_{1} k\left(x_{1}\right)+x_{2}^{2}\right)+\left( x_{1}+2 x_{2}\right) w\\
   &\quad  +(2k(x_1)+x_2+x_1)\theta_1
    +(x_1+2x_2)\theta_2\\
&\leqslant-\left(\left|x_{1}\right| \eta\left(\left|x_{1}\right|\right)+x_{2}^{2}\right)+\left(\left|x_{1}\right|+2\left|x_{2}\right|\right)|w|\\
&\qquad +(2|k(x_1)|+3|x_2|+2|x_1|)|\theta|\\
& \leqslant-\frac{1}{4} \left(\left|x_{1}\right| \eta\left(\left|x_{1}\right|\right)+x_{2}^{2}\right)
\end{align*}
if 
\begin{equation}\label{ineqnonlinearosc}
\begin{aligned}
   \left(\left|x_{1}\right| \eta\left(\left|x_{1}\right|\right)+x_{2}^{2}\right) &\geq\max\{2\left (\left|x_{1}\right|+2\left|x_{2}\right|\right)|w|,\\
   &\quad 4(2|k(x_1)|+3|x_2|+2|x_1|)|\theta|\}.
\end{aligned}
\end{equation}
 We will now prove that (\ref{eqn:exnewcond1}) implies (\ref{ineqnonlinearosc}):
We have already shown that
$|x|\geqslant\gamma(|w(t)|)$ implies $\left(\left|x_{1}\right| \eta\left(\left|x_{1}\right|\right)+x_{2}^{2}\right) \geqslant \{2\left (\left|x_{1}\right|+2\left|x_{2}\right|\right)|w|\}$ in the case where there is no time delay in the oscillator. Let us know prove that
\begin{align}\label{implicationineq1} 
   |x|\geqslant\gamma_{\theta}(|\theta(t)|)
\end{align}
implies 
\begin{align}\label{implicationineq2}
     \left(\left|x_{1}\right| \eta\left(\left|x_{1}\right|\right)+x_{2}^{2}\right) \geqslant4(2|k(x_1)|+3|x_2|+2|x_1|)|\theta|\}.
\end{align}
We consider two cases for this: $2|k(x_1)|+2|x_1|\geqslant3|x_2|$ and $2|k(x_1)|+2|x_1|< 3|x_2|$. We consider $k(x_1)=x_1^3$ here on wards. If $2|k(x_1)|+2|x_1|\geqslant3|x_2|$, then $|\theta|\leqslant\frac{\eta(|x_1|)}{8(2|x_1|^2+2)}\leqslant\eta(|x_1|)$, because otherwise we have a contradiction
 \begin{align*}
     |x|&=\sqrt{x_1^2+x_2^2}=\sqrt{x_1^2+\frac{4}{9}(x_1^3+x_1)^2}\\
     &\leqslant\max\{10|x_1|^4,10|x_1|^2\}\\
     &\leqslant\max\{10(\eta^{-1}(|\theta|))^4, 10(\eta^{-1}(|\theta|))^2\}.
 \end{align*}
Next we have, $|\theta|\leqslant\frac{\eta(|x_1|)}{8(2|x_1|^2+2)}$ and it follows that $4(2|x_1|^3+3|x_2|+2|x_1|)|\theta|\} \leqslant8(2|x_1|^2+2)|x_1|\frac{\eta(|x_1|)}{8(2|x_1|^2+2)}\leqslant|x_1|\eta(|x_1|)\leqslant\left(\left|x_{1}\right| \eta\left(\left|x_{1}\right|\right)+x_{2}^{2}\right)$.
In the opposite case, $2|x_1|^3+2|x_1|<3|x_2|$, then $|\theta|\leqslant\frac{|x_2|}{24}$, because otherwise we have a contradiction,
\begin{align*}
    |x|&=\sqrt{x_1^2+x_2^2}\leq\sqrt{(2x_1^3+2x_1)^2+x_2^2}\leqslant\sqrt{9x_2^2+x_2^2}\\
    &\leqslant\sqrt{10}|x_2|\leqslant24 \sqrt{10}|\theta|.
\end{align*}
Hence, we have $|\theta|\leqslant\frac{|x_2|}{24}$ and it follows that 
$4(2|x_1|^3+3|x_2|+2|x_1|)|\theta|\} \leqslant8(3|x_2|)|\theta|\leqslant|x_2|^2\leqslant(\left|x_{1}\right| \eta\left(\left|x_{1}\right|\right)+x_{2}^{2})$ as desired, which proves (\ref{implicationineq1}) implies (\ref{implicationineq2}). With $\theta(t)=(x_2(t-\tau)-x_2(t),-(x_2(t-\tau)-x_2(t)))^{\top}$ and  $x_2(t)$ being locally Lipschitz, we can further obtain the following using integral mean value theorem,
\begin{align*}
     |\theta(t)| &\leqslant \tau L|(-x_1^3(c)-x_2(c-\tau)+w(c))|\\
 &\leqslant \tau \max\{\gamma_1(|x_t|),\gamma_2(|w_t|)\},
\end{align*}
where $c \in (t-\tau,t)$ and $L$ is the Lipschitz constant corresponding to $k:[a,b]\to \mathbb{R}$ and $a,b \in \mathbb{R}$.
Define
\begin{align*}
 \gamma_1(r_1)&=\max_{|\lambda_1| \leqslant r_1,|\lambda_2| \leqslant r_1}10L|\lambda_1^3+\lambda_2|,\\
\gamma_2(r_2)&=10L|r_2|.
\end{align*}
We now define $\hat{\gamma} \in \mathcal{G}$ by $\hat{\gamma}\triangleq\max\{\gamma_{\theta}(\tau.\gamma_2(s)),\gamma(s)\}$, and following (\ref{eqn:exnewcond1}), we obtain
\begin{align*}
    |x(t)|_A &\geqslant \max\{\gamma_{\theta}(\tau.\gamma_1(|x_t|)),\hat{\gamma}(|w_t|)\}\\
    &\implies \dot{V}\leqslant -\frac{1}{4} \left(\left|x_{1}\right| \eta\left(\left|x_{1}\right|\right)+x_{2}^{2}\right).
\end{align*}
We can now choose $\tau^* >0$ such that for $\tau \in (0,\tau^*)$,
\begin{align*}
    \alpha_1^{-1}\circ\alpha_2 \circ \gamma_{\theta}(\tau.\gamma_1(s))<s, \quad \forall s \in (\chi,\Delta),
\end{align*}
holds, for each pair of strictly positive real numbers $\chi$, and $\Delta$. The analysis that follows assumes that $\tau$ can be further refined so that we achieve the same ISS gain as is for (\ref{nonlinearoscsys}), in accordance with \cite{teel1998connections} and Theorem \ref{robustness1}. The next example shows how theorems \ref{robustness2} and \ref{robustness1} differ by taking into account a compact, invariant set that contains the origin.
\subsection{Example 2}
Here, we focus on Stuart-Landau oscillator systems, which are used to model the behaviour of complex systems in a variety of contexts. For instance, they can be used to characterize electronic oscillators, semiconductor lasers, chemical reaction diffusion systems, and neuro-physiological phenomena. Studies on the stability and dynamic consensus of Stuart-Landau oscillators can be found in \cite{panteley2015stability} and \cite{panteley2020practical} respectively. A forced generalised Stuart-Landau oscillator, is described by  
\begin{equation}\label{forcedstuart}
\dot{z}=-\nu|z|^{2} z+\mu z+u,
\end{equation}
where $z \in \mathbb{C}$ represents the oscillator's state, $\nu, \mu \in \mathbb{C}$ are parameters defined as $\nu=\nu_{\mathrm{R}}+\mathrm{i} \nu_{I}$ and $\mu=\mu_{\mathrm{R}}+\mathrm{i} \mu_{I}$, respectively, and  $u \in \mathbb{C}$ is an input to the oscillator. Here, we take into account $\mu_R,\nu_R>0$. The compact set of equilibrium points for (\ref{forcedstuart}) is given by
\begin{align*}
A\triangleq \left\{z \in \mathbb{C}:|z|=\sqrt{\frac{\mu_{\mathrm{R}}}{\nu_{\mathrm{R}}}}\right\} \bigcup\{z=0\}.
\end{align*}
We define the norm $|\cdot|_A$, as follows.
$$
|z|_A=\left\{\begin{array}{ccc}
|z| & \text { if } & |z| \leqslant 0.7\sqrt{\alpha}, \\
\sqrt{\left.|| z\right|^{2}-\alpha \mid} & \text { if } & |z| > 0.7 \sqrt{\alpha}  \\
\alpha\triangleq \mu_{\mathrm{R}} / \nu_{\mathrm{R}}.
\end{array}\right.
$$
Let us consider the following candidate ISS-Lyapunov function
\begin{align*}
V(z)=\frac{1}{4 \nu_{\mathrm{R}}}\left[|z|^{2}-\alpha\right]^{2},
\end{align*}
where $\alpha=\mu_{\mathrm{R}} / \nu_{\mathrm{R}}$ and  $|z|^{2}=\bar{z} z$ and $\bar{z}$ signify the conjugate of $z$. Observe that $V(z)=0$ iff $z \in \mathcal{A}_{1}=\left\{z \in \mathbb{C}:|z|=\sqrt{\frac{\mu_{\mathrm{R}}}{\nu_{\mathrm{R}}}}\right\} $ and it is positive otherwise.\newline\newline
As obtained in \cite{panteley2015stability}, the time derivative of $V$ along the trajectories of (\ref{forcedstuart}) are
\begin{align}  \label{finalVdotstuart}
\dot{V}(z) \leqslant-\frac{1}{2}\left[|z|^{2}-\alpha\right]^{2}|z|^{2}+\frac{1}{2 \nu_{\mathrm{R}}^{2}}|u|^{2}. 
\end{align}
The right-hand side of the inequality mentioned above is then bounded in terms of $|z|_A$. To bound the term $\left[|z|^{2}-\alpha\right]^{2}|z|^{2}$ in the  two cases ($|z|_A\leqslant 0.7\sqrt{\alpha}$ and $|z|_A>0.7\sqrt{\alpha}$) independently. Denoting $c_{3}(|z|_A)\triangleq\min \left\{0.51 \alpha^{2}|z|_A^2, 0.49 \alpha\right|z|_A^4\}$ and combining the two cases together, we obtain
\begin{align*}
\dot{V}(z) &\leqslant -c_{3}(|z|_A)+\frac{1}{2 \nu_{\mathrm{R}}^{2}}|u|^{2},
\end{align*}
which can also be rewritten as
\begin{align*}
    |z|_A\geqslant c_3^{-1}\big(\frac{4}{ \nu_{\mathrm{R}}^{2}}|u|^{2}\big)\triangleq \gamma(|u|) \implies \dot{V} \leqslant -\frac{7}{8}c_3(|z|_A).
\end{align*}
Thus, the system (\ref{forcedstuart}) is set ISS with gain $\gamma \in \mathcal{K}_{\infty}$.\newline\newline
We now add input delay perturbation, which is $g(z_t)=\nu|z(t)|^{2} z(t)-\nu|z(t-\tau)|^{2} z(t-\tau)-\mu z(t)+\mu z(t-\tau)$. As a result, the time delay modification of (\ref{forcedstuart}) is as follows:
\begin{align*}
    \dot{z}(t)&=-\nu|z(t-\tau)|^{2} z(t-\tau)+\mu z(t-\tau)+u(t).
\end{align*}
We now employ the mean value theorem to obtain
\begin{align*}
    |g(z_t)|&\leq\tau|\mu(-\nu|z(\phi-\tau)|^{2} z(\phi-\tau)\\
    &\qquad \qquad +\mu z(\phi-\tau)+u(\phi))|\\
    &+\tau|3|z(\phi)|^2\nu(-\nu|z(\phi-\tau)|^{2} z(\phi-\tau)\\
    &\qquad +\mu z(\phi-\tau)+u(\phi))|.
\end{align*}
It is difficult to establish the requirement that there exists a $k \in\mathcal{K}_{\infty}$, such that $|g(z_t)|\leqslant k(|V_t|)$, even in situations where Theorem \ref{robustness2} is applicable. This is because we are unaware of any shared relationships between $u(\phi)$ and $z(\phi),z(\phi-\tau)$, where $\phi\in[t-\tau,t]$. \newline\newline
We propose that the set ISS of the time delayed system can be demonstrated using Theorem \ref{robustness1} rather than having to go through the process of validating the above requirements of Theorem \ref{robustness2}. In accordance with (\ref{forcedstuart}), we take into consideration the following system
\begin{align}\label{stuartourapproach}
\dot{z}=w+u,
\end{align}
where $u\in\mathbb{C}$. We take into account the feedback $w=k(z)=-\nu|z|^{2} z+\mu z$ with the parameters $\nu=\nu_{\mathrm{R}}+\mathrm{i} \nu_{I}$ and $\mu=\mu_{\mathrm{R}}+\mathrm{i} \mu_{I}$, and $\mu_R,\nu_R>0$. This system (\ref{stuartourapproach}) is set ISS with respect to $A$. Now, we'll examine whether the system (\ref{stuartourapproach}) is set ISS with time-delay in feedback. To do the same, we first demonstrate that the subsequent system
\begin{equation}\label{forcedstuartwiththeta}
\dot{z}=-\nu|z|^{2} z+\mu z+u+\theta,
\end{equation}
with 
\begin{align*}
    \theta &=\nu|z(t)|^{2} z(t)-\nu|z(t-\tau)|^{2} z(t-\tau)\\
    &\qquad -\mu z(t)+\mu z(t-\tau),
\end{align*}
satisfies $\dot{V}\leqslant -\frac{1}{2}\left[|z|^{2}-\alpha\right]^{2}|z|^{2}+\frac{1}{ \nu_{\mathrm{R}}^{2}}|u|^{2}+\frac{1}{ \nu_{\mathrm{R}}^{2}}|\theta|^{2}$, or
\begin{align} \nonumber
    |z|_A&\geqslant \max\big\{c_3^{-1}\big(\frac{2}{ \nu_{\mathrm{R}}^{2}}|\theta|^{2}\big),c_3^{-1}\big(\frac{4}{ \nu_{\mathrm{R}}^{2}}|u|^{2}\big)\big\}\\ \label{vdotineqstuart}
    &\implies \dot{V} \leqslant -\frac{1}{4}c_3(|z|_A).
\end{align}
This is inferred from (\ref{finalVdotstuart}), when the time derivative of the ISS-Lyapunov function is taken along the trajectories (\ref{forcedstuartwiththeta}). It is important to note that we were unable to determine $\gamma_{\theta}(|\theta|)$, which yields $\dot{V}\leqslant -\frac{7}{16}c_3(|z|_A)$. However, this won't have an impact on our analysis as we have some definite negative function in $|z|_A$.
$\theta(t)= k(z(t-\tau))-k(z(t))$ and $k$ is a locally Lipschitz function. Thus, we obtain
\begin{align*}
    |\theta(t)| &\leqslant L\Big|\int_{t-\tau}^{t}(-\nu|z(s-\tau)|^{2} z(s-\tau)\\
    &\qquad \qquad +\mu z(s-\tau)+u(s))ds\Big|.
\end{align*}
for a particular Lipschitz constant when the function $k$ is specified in a local domain. The results of applying the integral mean value theorem to the aforementioned equation are as follows:
\begin{align*}
     |\theta(t)| &\leqslant L \tau |-\nu|z(c-\tau)|^{2} z(c-\tau)+\mu z(c-\tau)+u(c)|,
\end{align*}
where $c \in (t-\tau,t)$, and
\begin{align*}
     &L\tau|-\nu|z(c-\tau)|^{2} z(c-\tau)+\mu z(c-\tau)+u(c)|\\
     &\leqslant \tau\max\{\gamma_1(|z_t|_A),\gamma_2(|u_t|)\}.
\end{align*}
Define
\begin{align*}
 \gamma_1(r_1)&=\max_{|\lambda|_A \leqslant r_1}10L|-\nu|\lambda|^{2} \lambda+\mu \lambda|,\\
\gamma_2(r_2)&=10L|r_2|.
\end{align*}
We now define $\hat{\gamma} \in \mathcal{G}$ by $\hat{\gamma}\triangleq\max\{\gamma_{\theta}(\tau.\gamma_2(s)),\gamma(s)\}$, where $\gamma_{\theta}(s)=c_3^{-1}\big(\frac{2}{ \nu_{\mathrm{R}}^{2}}|s|^{2}\big)$. As a result of (\ref{vdotineqstuart}),
\begin{align*}
    |z(t)|_A &\geqslant \max\{\gamma_{\theta}(\tau.\gamma_1(|z_t|_A)),\hat{\gamma}(|u_t|)\}\\
    &\implies \dot{V}\leqslant -\frac{c_3(|z|_A)}{4}.
\end{align*}
is implied.
We now perform the identical procedures as in Subsection \ref{exam1} to prove the findings from the Theorem \ref{robustness1} and to establish the set ISS of (\ref{forcedstuartwiththeta}).

\section{Conclusion}\label{conclusion}
A Razumikhin-type set stability theorem has been interpreted in terms of the ISS nonlinear small-gain theorem and extended to cover the case of persistent disturbances. Set input-to-state stabilizability (and set-global asymptotic stabilizability as a special case) is shown to be robust, in an appropriate sense, to small time delays at the input for nonlinear control systems nominally described by ordinary differential equations. We also compare our result to a related result which shows robustness of set ISS to time delay using illustrative examples.

\bibliographystyle{IEEEtran}
\bibliography{ref}

\end{document}